# Integrated Security Mechanisms for Weight Protection in Memristive Crossbar Arrays

Muhammad Faheemur Rahman, Wayne Burleson
Department of Electrical and Computer Engineering, University of Massachusetts Amherst
Email: faheemur@umass.edu, burleson@umass.edu

*Abstract*— Memristive crossbar arrays enable in-memory computing by performing parallel analog computations directly within memory, making them well-suited for machine learning, neural networks, and neuromorphic systems. However, despite their advantages, non-volatile memristors are vulnerable to security threats (such as adversarial extraction of stored weights when the hardware is compromised. Protecting these weights is essential since they represent valuable intellectual property resulting from lengthy and costly training processes using large, often proprietary, datasets. As a solution we propose two security mechanisms: Keyed Permutor and Watermark Protection Columns; where both safeguard critical weights and establish verifiable ownership (even in cases of data leakage). Our approach integrates efficiently with existing memristive crossbar architectures without significant design modifications. Simulations across 45nm, 22nm, and 7nm CMOS nodes, using a realistic interconnect model and a large RF dataset, show that both mechanisms offer robust protection with under 10% overhead in area, delay and power. We also present initial experiments employing the widely known MNIST dataset; further highlighting the feasibility of securing memristive in-memory computing systems with minimal performance trade-offs.

## I. Introduction

The rise of artificial intelligence (AI) and machine learning (ML) has exposed limitations in traditional computing architectures, particularly the memory bottleneck caused by separating memory and processing units [1]. In-Memory Computing reduces this issue by enabling data processing directly within memory, significantly improving energy efficiency and computational throughput [2]. Among various hardware platforms, memristive crossbar arrays have emerged as a promising solution due to their non-volatility and ability to perform analog matrix-vector multiplications efficiently in a single step (outperforming traditional digital methods) [3].

Despite these advantages, the use of memristive crossbars in processing real-world, complex datasets, such as those from radio frequency (RF) signal environments remains relatively unexplored [4]. Integrating such datasets introduces new challenges, especially in ensuring the security of stored weights. Due to the persistent nature of memristor states, sensitive model parameters, such as weights, may be exposed to adversaries, making the system vulnerable to intellectual property theft and malicious tampering [4]. To address these challenges, we propose and implement two security mechanisms: (1) Keyed Permutor and (2) Watermark Protection Columns; and evaluate their power, performance, and area (PPA) overheads relative to an unsecured baseline. We also design and simulate both large and small memristive crossbar arrays to compare the effectiveness of different security configurations. Moreover, a complex RF dataset is incorporated into the secured architecture to demonstrate its applicability to advanced signal processing tasks. Furthermore, the impact of interconnect wire modeling is analyzed across multiple CMOS technology nodes, offering insights into the scalability of future hardware implementations.

## II. Proposed Approaches

We consider a worst-case white-box scenario where an adversary with advanced tools extracts sensitive data, such as network architecture and crossbar weights. Even without direct cell access, weights can be inferred through peripheral circuitry and cloned, replicating the model without costly training. To counter this, we introduce two security mechanisms below that protect the integrity and ownership of the crossbar-based array.

**Keyed Permutor:** The Keyed Permutor requires a secret key to remap input signals to memristor rows in an unpredictable manner, obscuring the true locations of stored weights. Unlike direct row activation, the key-controlled permutation ensures that selecting a specific input could trigger any of several rows (e.g., Row 1 may activate Row 5, 1000, or remain unchanged), making physical observation ineffective without the key. To further enhance security, our design employs triplet swaps, increasing permutation complexity and making brute-force attacks infeasible. The triplet swap offers strong security with a large key space (approximately $2^{109}$) while maintaining reasonable hardware overhead (only 2.34% transistor increase for a 128×128 array). For system integrity, keys must be securely stored and periodically updated to prevent compromise.

**Watermark Protection Columns:** Attackers may attempt to tamper with or obscure a watermark to claim ownership or prevent verification [5]. We add two dedicated columns for watermark checking, which store no neural weights and are placed at the array's end to model a worst-case scenario. However, their position can vary, for example, at the beginning, end, or spread out to avoid detection. The watermark columns are initialized with fixed patterns (e.g., predefined voltage levels or resistance states) that do not interfere with normal computation. During inference, these known patterns are verified through their distinct current signatures to verify watermark integrity. Moreover, to prevent attackers from identifying them, the watermark columns are designed to mimic regular columns in structure and behavior, with variable placement and dummy activity to blend into the array.

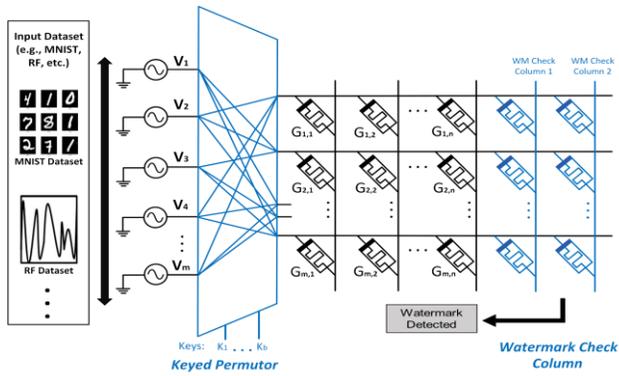

Fig. 1. Memristive Crossbar Array with two integrated security mechanisms

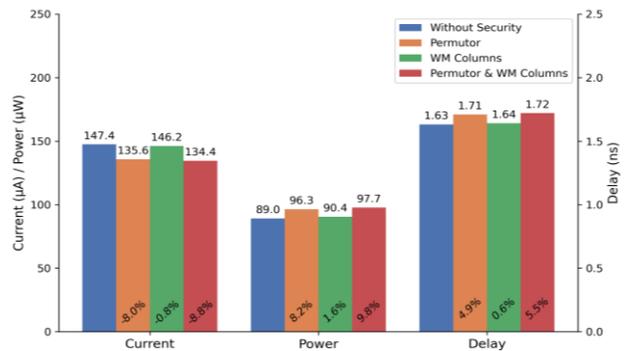

Fig. 2. Comparison of Current, Power, and Delay

## III. Experimental Setup & Methodology

We first used the MNIST dataset, downsampled and flattened to match the input row size of the memristive crossbar array. After normalization, the data was fed into the array for matrix-vector multiplication. We then applied a more complex RF dataset denoted as "Long Range" captured from devices that use LoRa technology. In both cases, down-sampling is necessary to fit large datasets to the array dimensions. In the crossbar array, inputs include normalized analog voltages from datasets, and outputs are analog currents from columns after matrix-vector multiplication. The internal state refers to memristor conductance values that represent stored weights. In our current evaluation, the weights were not learned through training but were predefined for structural and power-delay simulation purposes, focusing on security overhead and circuit behavior rather than classification accuracy.

To support secure computation, we adopted the 1T1R (one transistor-one memristor) cell structure, which prevents sneak path currents and ensures accurate weight tuning [3]. A interconnect model was also integrated to account for parasitic effects at the nanoscale, improving simulation realism. We conducted simulations using HSPICE across three CMOS technology nodes: the 45nm foundry node, 22nm Bulk PTM, and 7nm FinFET PTM. Furthermore, arrays of different sizes: 10×10, 128×10, and 256×128 were evaluated. Fig. 1 demonstrates the application of our two security mechanisms. We first simulated the baseline arrays, then added each security mechanism individually, and finally combined both. For each case, we measured power, delay, and column current, and calculated the overhead relative to the unsecured design.

## IV. Discussion And Conclusion

Our memristive crossbar array successfully performs analog matrix-vector multiplication based on Kirchhoff's and Ohm's laws, confirming its functional accuracy. The proposed Keyed Permutor and Watermark Protection Columns integrate effectively into existing architectures with minimal design changes. Fig. 2 highlights simulation results for a 256×128 array at the 45 nm node, showing modest overhead: an 8.8% drop in column current, a 5.5% increase in delay, and a 9.8% increase in power. Transistor count rises by just 2.34% using the triplet-swap configuration. These trends hold consistently across smaller arrays (128×10, 10×10) and advanced nodes (22 nm, 7 nm), demonstrating scalability and robustness.

With the cost of training major ML models projected to exceed $500M by 2030 [4], efficient and secure in-memory architectures like memristive crossbars are increasingly vital. Their low power, high speed, and density make them ideal for next-generation computing, including neuromorphic systems [3]. As adoption grows, securing stored weights becomes critical. Our proposed mechanisms obscure the mapping between inputs and stored values, resist tampering, and support ownership verification with low overhead. These results show that security can be implemented effectively with negligible cost, demonstrating that protection does not need to come at the expense of performance or efficiency. Future work will focus on Process, Voltage, and Temperature analysis, performing Monte Carlo simulations on key design parameters, implementing the proposed security mechanisms on larger crossbar arrays, and further optimizing power and delay performance.

### Acknowledgment

The research was sponsored by the Army Research Laboratory and was accomplished under Cooperative Agreement Number W911NF-23-2-0014. The views and conclusions contained in this document are those of the authors and should not be interpreted as representing the official policies, either expressed or implied, of the Army Research Laboratory or the U.S. Government. The U.S. Government is authorized to reproduce and distribute reprints for Government purposes notwithstanding any copyright notation herein.